\documentclass{article}

\usepackage{spconf,amsmath,graphicx}
\usepackage{amssymb,amsfonts}
\usepackage[hyphens]{url}
\usepackage{bm,cite,mathtools}
\usepackage{subfig}
\usepackage{mathrsfs}

\usepackage{bbm}
\usepackage{float}
\usepackage{parselines}

\newcommand{\argmin}{\mathop{\rm arg~min}\limits}

\title{Kernel interpolation of acoustic transfer functions\\with adaptive kernel for directed and residual reverberations}
\name{Juliano G. C. Ribeiro, Shoichi Koyama, and Hiroshi Saruwatari}
\address{The University of Tokyo, 7-3-1 Hongo, Bunkyo-ku, Tokyo 113-8656, Japan}

\begin{document}
\ninept
\maketitle
\begin{abstract}
An interpolation method for region-to-region acoustic transfer functions (ATFs) based on kernel ridge regression with an adaptive kernel is proposed. Most current ATF interpolation methods do not incorporate the acoustic properties for which measurements are performed. Our proposed method is based on a separate adaptation of directional weighting functions to directed and residual reverberations, which are used for adapting kernel functions. Thus, the proposed method can not only impose constraints on fundamental acoustic properties, but can also adapt to the acoustic environment. Numerical experimental results indicated that our proposed method outperforms the current methods in terms of interpolation accuracy, especially at high frequencies. 

\end{abstract}
\section{Introduction}
\label{sec:intro}

An acoustic environment is characterized by its acoustic impulse response (AIR) in the time domain or acoustic transfer function (ATF) in the frequency domain between the source and the receiver. Understanding the acoustic characteristics of the target environment is important because they have extremely broad applications. Our focus is the ATF interpolation problem from measurements at discrete positions. 

Most of the current methods of ATF and AIR approximation are derived for point-to-point or point-to-region cases; e.g., modeling ATFs as linear time-invariant systems with poles and/or zeros~\cite{IIR_ATF,Haneda:IEEE_J_SAP1999}, modeling of the AIR using sparse expansions of elementary wave components~\cite{Mignot:IEEE_ACM_J_ASLP2014, Antonello:TASLP2017, Das:ICASSP2021}, and AIR reconstruction by deep learning using training data~\cite{Pezzoli:Sensors2022}. However, these methods do not account for source variation and do not adopt a generalized framework for the ATF that can be applied freely in most environments.



The region-to-region ATF interpolation, i.e., interpolating ATFs between regions of sources and receivers, will further expand its applications. Samarasinghe~et al. proposed a method based on basis expansion to a finite set of spherical wave functions~\cite{Samarasinghe}. We proposed a method based on kernel ridge regression (KRR)~\cite{Ribeiro:IEEE_SAM2020,Ribeiro:IEEE_ACM_J_ASLP2022}, which can be regarded as an extension of the method in \cite{Samarasinghe} to infinite-dimensional expansion. Constraints on fundamental acoustic properties, such as the Helmholtz equation and reciprocity, are imposed by designing the kernel function based on plane-wave expansion. This method is later extended by incorporating a weighting function to prioritize the directions that ATF will have a large amplitude~\cite{Ribeiro:ICASSP2022}. 



Current region-to-region ATF interpolation methods do not incorporate the acoustic properties for which measurements are performed. Several attempts on incorporating the properties of an acoustic environment have been made in the context of sound field estimation from multiple microphone measurements~\cite{Horiuchi:WASPAA2021,Shigemi:IWAENC2022}. We propose a region-to-region ATF interpolation method adapted to the acoustic environment with constraining the fundamental acoustic properties. The proposed method is based on the separate modeling of directed and residual sound fields. The weighting function for the directed field is formulated by the superposition of unimodal functions, and for the residual field is modeled by using neural networks. Although the hyperparameters for the proposed model are obtained by a gradient-descent-based optimization, the estimate is still obtained by kernel ridge regression in a closed form. We evaluate our proposed method by numerical experiments, and compared it with current ATF interpolation methods.

\section{Problem statement and prior works}
\label{sec:prelim}

\begin{figure}[!t]
\centering
\centerline{\includegraphics[width=0.65\columnwidth]{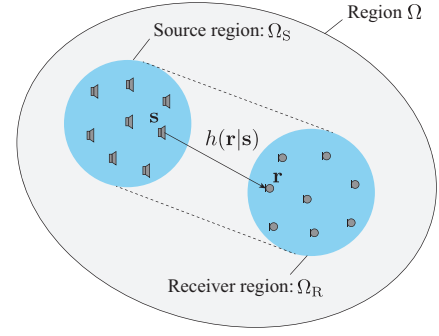}}
  \caption{Diagram exemplifying a region-to-region ATF interpolation problem.}
  \label{fig:prob}
\end{figure}

Suppose a region of arbitrary geometry $\Omega \subseteq \mathbb{R}^3$ with stationary acoustic properties. We further define two regions within $\Omega$: a source region $\Omega_\mathrm{S}\subset \Omega$ and a receiver region $\Omega_\mathrm{R}\subset \Omega$. Our objective is to estimate the ATF $h: \Omega_\mathrm{R} \times \Omega_\mathrm{S} \to \mathbb{C}$ between any source/receiver pairs of positions $\mathbf{s}\in \Omega_\mathrm{S}$ and $\mathbf{r} \in \Omega_\mathrm{R}$ (see Fig.~\ref{fig:prob}).


\subsection{Region-to-region ATF interpolation problem}

First, the region-to-region ATF interpolation problem is mathematically defined in the frequency domain, as described in \cite{Samarasinghe,Ribeiro:IEEE_ACM_J_ASLP2022}. The ATF $h$ is assumed to be the superposition of a known direct component $h_\mathrm{D}$, represented by the free-field Green's function $G_0$, and an unknown reverberant component $h_\mathrm{R}$, satisfying the homogeneous Helmholtz equation for $\mathbf{r}$ and $\mathbf{s}$:
\begin{align}
& h(\mathbf{r}|\mathbf{s}) = h_\mathrm{D}(\mathbf{r}|\mathbf{s}) + h_\mathrm{R}(\mathbf{r}|\mathbf{s})\\ 
&h_\mathrm{D}(\mathbf{r}|\mathbf{s}) = G_0(\mathbf{r}|\mathbf{s}) = \frac{\mathrm{e}^{\mathrm{i}k\|\mathbf{r} - \mathbf{s}\|}}{4\pi \|\mathbf{r} - \mathbf{s}\|}\\
&(\nabla^2_\mathbf{r}+k^2)h_\mathrm{R}(\mathbf{r}|\mathbf{s}) =(\nabla^2_\mathbf{s}+k^2)h_\mathrm{R}(\mathbf{r}|\mathbf{s}) =0,
\end{align}
where $\mathbf{r}|\mathbf{s}$ represents the source/receiver pair, $\mathrm{i}$ is the imaginary unit, and $\nabla^2_{\mathbf{r}}$ and $\nabla^2_\mathbf{s}$ are the Laplacian operators in relation to $\mathbf{r}$ and $\mathbf{s}$, respectively. 

$M$ receivers and $L$ sources are distributed at $\{\mathbf{s}_l\}_{l=1}^L\subset \Omega_\mathrm{S}$ and $\{\mathbf{r}_m\}_{m=1}^M\subset \Omega_\mathrm{R}$, respectively. Given $N$ ($=ML$) ATF measurements between all $\mathbf{q}_{n} = \mathbf{r}_m|\mathbf{s}_l$ possible pairs with index $n=m+(l-1)M$ ($\in\{1,\ldots,N\}$), our objective is to find $\hat{h}_\mathrm{R}$ by solving
\begin{equation}
 \hat{h}_{\mathrm{R}} = \argmin_{g\in\mathscr{H}} \sum_{n=1}^N |y_n-g(\mathbf{q}_n)|^2+\lambda \|g\|_{\mathscr{H}}^2,
 \label{eq:optim}
\end{equation}
where $g$ is a generic function, $\{y_n\}_{n=1}^N$ are the reverberant component measurements $\{h_\mathrm{R}(\mathbf{q}_n)\}_{n=1}^N$ with added noise, $\lambda>0$ is a regularization constant, and $\mathscr{H}$ is the feature space of the estimation. Then, the estimate of ATF $\hat{h}$ is obtained by adding $h_\mathrm{D}$ as
\begin{equation}
\hat{h}(\mathbf{r}|\mathbf{s}) = \hat{h}_\mathrm{R}(\mathbf{r}|\mathbf{s})  + h_\mathrm{D}(\mathbf{r}|\mathbf{s}).
\end{equation}


\subsection{Kernel ridge regression with directional weighting for ATF interpolation}

In \cite{Ribeiro:ICASSP2022,Ribeiro:IEEE_ACM_J_ASLP2022}, the feature space $\mathscr{H}$ is assumed to be a reproducing kernel Hilbert space (RKHS) constraining fundamental properties of the acoustic field, thereby satisfying the homogeneous Helmholtz equation and reciprocity. To define $\mathscr{H}$, we model the reverberant component as a form of plane wave expansion (or \textit{Herglotz wave function})~\cite{Colton2003, Ikehata:HMJ2005}:
\begin{align}
h_\mathrm{R}(\mathbf{r}|\mathbf{s}) &= \mathcal{T}(\tilde{h}_\mathrm{R};\mathbf{r}|\mathbf{s}) \notag\\
&:= \int_{\mathbb{S}^2\times \mathbb{S}^2} \mathrm{e}^{\mathrm{i}k(\hat{\mathbf{r}}\cdot \mathbf{r} + \hat{\mathbf{s}}\cdot \mathbf{s}) } \tilde{h}_{\mathrm{R}}(\hat{\mathbf{r}}, \hat{\mathbf{s}}) \mathrm{d}\hat{\mathbf{r}} \mathrm{d} \hat{\mathbf{s}},
\end{align}
where $\mathbb{S}^2$ is the unit sphere in $\mathbb{R}^3$, $\hat{\mathbf{s}}\in \mathbb{S}^2$ and $\hat{\mathbf{r}}\in \mathbb{S}^2$ represent plane wave component directions relating to source and receiver, respectively, and $\tilde{h}_{\mathrm{R}}$ is the plane-wave weighting function. Thus, the inner-product space $(\mathscr{H}, \langle \cdot, \cdot \rangle_{\mathscr{H}})$ is defined as~\cite{Ribeiro:ICASSP2022}
\begin{align}
&\mathscr{H} := \left\{ \mathcal{T}(\tilde{h}_{\mathrm{R}};\mathbf{r}|\mathbf{s}):\tilde{h}_{\mathrm{R}}\in L^2(\mathbb{S}^2\times \mathbb{S}^2, w), \right. \notag \\
&\hspace{80pt} \ \left. \tilde{h}_{\mathrm{R}}(\hat{\mathbf{r}}, \hat{\mathbf{s}}) = \tilde{h}_{\mathrm{R}}(\hat{\mathbf{s}}, \hat{\mathbf{r}}), \ \forall \hat{\mathbf{r}}, \hat{\mathbf{s}}\in \mathbb{S}^2 \right\}\\
&\langle h_{\mathrm{R},1}, h_{\mathrm{R},2}\rangle_{\mathscr{H}} := \int_{\mathbb{S}^2\times \mathbb{S}^2} \frac{\overline{\tilde{h}_{\mathrm{R},1}(\hat{\mathbf{r}}, \hat{\mathbf{s}})} \tilde{h}_{\mathrm{R},2}(\hat{\mathbf{r}}, \hat{\mathbf{s}})}{w(\hat{\mathbf{r}}, \hat{\mathbf{s}})}\mathrm{d}\hat{\mathbf{r}}\mathrm{d} \hat{\mathbf{s}},
\end{align}
where $\overline{\cdot}$ is the complex conjugate operator, $w:\mathbb{S}^2\times \mathbb{S}^2\rightarrow \mathbb{R}_+$ is the directional weighting function, and $L^2(\mathbb{S}^2\times \mathbb{S}^2, w)$ is the space of square integrable functions. The equality $\tilde{h}_{\mathrm{R}}(\hat{\mathbf{r}}, \hat{\mathbf{s}}) = \tilde{h}_{\mathrm{R}}(\hat{\mathbf{s}}, \hat{\mathbf{r}})$ represents the reciprocity of ATF. Thus, $(\mathscr{H}, \langle \cdot, \cdot \rangle_{\mathscr{H}})$ is the Hilbert space with the reproducing kernel function $\kappa$ given by
\begin{equation}
\kappa(\mathbf{r}|\mathbf{s}, \mathbf{r}^\prime|\mathbf{s}^\prime) = \mathcal{T}\left (w(\hat{\mathbf{r}}, \hat{\mathbf{s}})\frac{\mathrm{e}^{-\mathrm{i}k(\hat{\mathbf{r}}\cdot \mathbf{r}^\prime + \hat{\mathbf{s}}\cdot \mathbf{s}^\prime)} + \mathrm{e}^{-\mathrm{i}k(\hat{\mathbf{r}}\cdot \mathbf{s}^\prime + \hat{\mathbf{s}}\cdot \mathbf{r}^\prime)} }{2}; \mathbf{r}|\mathbf{s} \right )\label{eq:kernel_func},
\end{equation}
where $\mathbf{r}^\prime\in \Omega_\mathrm{R}$ and $\mathbf{s}^\prime \in \Omega_\mathrm{S}$.

Since $\mathscr{H}$ is a RKHS, \eqref{eq:optim} has a closed-form solution as KRR~\cite{Murphy:ProbML}:
\begin{equation}
\hat{h}_\mathrm{R}(\mathbf{r}|\mathbf{s}; Q, \mathbf{y}, w) = \bm{\kappa} (\mathbf{r}|\mathbf{s}) (\mathbf{K}+\lambda \mathbf{I})^{-1}\mathbf{y},
\label{eq:krr}
\end{equation}
where $Q=\{\mathbf{q}_1,\dots,\mathbf{q}_N\}$ is the set of source/receiver pairs, $\mathbf{y}=[y_1,\ldots,y_N]^{\mathsf{T}}$ is the measurement vector, $\mathbf{I}$ is the identity matrix, $\bm{\kappa}(\mathbf{r}|\mathbf{s}) = [ \kappa(\mathbf{r}|\mathbf{s}, \mathbf{q}_1), \dots, \kappa(\mathbf{r}|\mathbf{s}, \mathbf{q}_N)]$ is the kernel function vector, and $\mathbf{K} = [\kappa(\mathbf{q}_n, \mathbf{q}_{n^\prime})]_{n,n^\prime \in \{1,2,\dots,N\}}$ is the Gram matrix.



\subsection{Directional weighting by sunken sphere}

The above RKHS has a freedom of design for the weighting function $w(\hat{\mathbf{r}}, \hat{\mathbf{s}})$. When there is no prior information on the directionality, this weighting function should be set as uniform, i.e., $w(\hat{\mathbf{r}}, \hat{\mathbf{s}})=1/16\pi^2$~\cite{Ribeiro:IEEE_ACM_J_ASLP2022}. In our previous study~\cite{Ribeiro:ICASSP2022}, the weighting function is defined so that the gain of the direct component is minimized since the direct component is removed from the measurements. 

We consider that the weight $w(\hat{\mathbf{r}}, \hat{\mathbf{s}})$ is separable for $\hat{\mathbf{r}}$ and $\hat{\mathbf{s}}$, that is,
\begin{equation}
w(\hat{\mathbf{r}}, \hat{\mathbf{s}}) = \varphi(\hat{\mathbf{r}})\varphi(\hat{\mathbf{s}}).
\end{equation}
Then, $\varphi(\hat{\mathbf{v}})$, where $\hat{\mathbf{v}}\in \mathbb{S}^2$ is $\hat{\mathbf{r}}$ or $\hat{\mathbf{s}}$, is defined as a fixed function having a shape of a sunken sphere as
\begin{equation}
\varphi(\hat{\mathbf{v}}) = \frac{1}{4\pi} \left ( 1+\gamma^2-\frac{\cosh(\zeta \hat{\mathbf{v}} \cdot \hat{\mathbf{v}}_0)}{\cosh(\zeta)} \right ),
\label{eq:weight_sunkensphere}
\end{equation}
where $\zeta,\gamma \in \mathbb{R}_+$ are hyperparameters, and $\hat{\mathbf{v}}_0\in \mathbb{S}^2$ is the direction connecting the centers of the regions. Thus, the directional gain of $\hat{\mathbf{v}}_0$ is minimized by using this weighting function.

\section{Adaptive Kernel for Directed and Residual Reverberations}



The ATF interpolation method presented in Sect.~\ref{sec:prelim} can estimate the ATF based on KRR with constraints of the fundamental acoustic properties and directionality defined in \eqref{eq:weight_sunkensphere}. However, properties of the acoustic environment in which the ATF measurements are obtained cannot be incorporated. We consider adapting the directional weighting function $w(\hat{\mathbf{r}}, \hat{\mathbf{s}})$ to the acoustic environment from $\mathbf{y}$. 
Thus, the ATF can be interpolated in a region-to-region manner by KRR \eqref{eq:krr} with adaptive kernel under the constraint of fundamental acoustic properties. 


We consider a separate model of directed and residual reverberations. Plane waves that reflect from walls generally have strong directionality; therefore, $w$ should have large gain for these components. The directivity representing the remaining residual field will have small gain in general and be more complex. The models for directed and residual fields and an optimization method for their hyperparameters are explained in the subsequent sections. The weighting functions for directed and residual field are denoted by $w_{\mathrm{dir}}$ and $w_{\mathrm{res}}$, respectively, and add up together to $w$. 



\subsection{Directed reverberation model}

It is necessary for the directed field model to represent several sound waves from the boundary $\Omega$. First, we assume that $w_{\mathrm{dir}}$ is separable for $\hat{\mathbf{r}}$ and $\hat{\mathbf{s}}$. 
\begin{align}
w_{\mathrm{dir}}(\hat{\mathbf{r}}, \hat{\mathbf{s}}) = \varphi_{\mathrm{dir}}(\hat{\mathbf{r}}) \varphi_{\mathrm{dir}}(\hat{\mathbf{s}})
\end{align}
Then, we represent each $\varphi_{\mathrm{dir}}$ as a convex combination of the unimodal weighting functions derived from the von Mises--Fisher distribution~\cite{Mardia:Directional_Statistics} as
\begin{align}
&\varphi_{\mathrm{dir}}(\hat{\mathbf{v}}) = \sum_{d=1}^D \alpha_d \frac{\mathrm{e}^{\beta_d \hat{\mathbf{v}}\cdot \hat{\mathbf{v}}_d}}{4\pi C(\beta_d)}, \label{eq:weight_early_phi_w}\\
&\|\bm{\alpha}\|_1 = 1,\\
&C(\beta_d) = 
\begin{cases}
\frac{\sinh(\beta_d)}{\beta_d}, & \beta_d \neq 0 \\ 
1, & \beta_d = 0
\end{cases},
\end{align}
where $D$ is the number of unimodal functions, $\hat{\mathbf{v}}_d$ and $\beta_d$ represent direction and spreading of the $d$th unimodal function, respectively, $\bm{\alpha} = [\alpha_1, \ldots, \alpha_D]^\mathsf{T}\subset{\mathbb{R}_+^D}$ is the vector of the weight coefficients, and $C$ is the normalization functions. $\{\beta_d\}_{d=1}^D$ is summarized as $\bm{\beta}=[\beta_1,\ldots,\beta_D]^\mathsf{T}$. This model is also used in multiple kernel learning for sound field estimation~\cite{Horiuchi:WASPAA2021}.

The kernel function $\kappa_\mathrm{dir}$ defined with the weighting function $w_{\mathrm{dir}}$ can be obtained in a closed-form as
\begin{align}
&\kappa_\mathrm{dir}(\mathbf{r}|\mathbf{s}, \mathbf{r}^\prime|\mathbf{s}^\prime;w_\mathrm{dir})\notag \\
&=\frac{1}{2} (\kappa_{\varphi_{\mathrm{dir}}}(\mathbf{r}, \mathbf{r}^\prime;\varphi_\mathrm{dir})\kappa_{\varphi_{\mathrm{dir}}}(\mathbf{s}, \mathbf{s}^\prime;\varphi_\mathrm{dir})\notag \\
&\hspace{20pt}+\kappa_{\varphi_{\mathrm{dir}}}(\mathbf{s}, \mathbf{r}^\prime;\varphi_\mathrm{dir})\kappa_{\varphi_{\mathrm{dir}}}(\mathbf{r}, \mathbf{s}^\prime;\varphi_\mathrm{dir}))\\
&\kappa_{\varphi_{\mathrm{dir}}}(\mathbf{r}, \mathbf{r}^\prime;\varphi_{\mathrm{dir}}) := \sum_{d=1}^D \alpha_d \frac{j_0(\eta(k(\mathbf{r} - \mathbf{r}^\prime) -\mathrm{i}\beta_d \hat{\mathbf{v}}_d))}{C(\beta_d)},\\
&\eta(\mathbf{z}) := \sqrt{\mathbf{z}^\mathsf{T} \mathbf{z}},\ \mathbf{z}\in\mathbb{C}^3.
\end{align}
The directions $\{\hat{\mathbf{v}}_d\}_{d=1}^D$ should be uniformly distributed on $\mathbb{S}^2$. By setting $\hat{\mathbf{v}}_1$ to $\hat{\mathbf{v}}_0$ and $\beta_1$ to $0$, $w_{\mathrm{dir}}$ is forced not to prioritize the direction $\hat{\mathbf{v}}_0$ in $w_{\mathrm{dir}}$.

\subsection{Residual reverberation model}

The residual field generally has smaller amplitudes than the directed, and the weighting function $w_{\mathrm{res}}$ should have a complex shape. Since it is difficult to represent $w_{\mathrm{res}}$ with simple analytic functions, we use neural networks to model $w_{\mathrm{res}}$ with parameters $\bm{\theta}$: $w_{\mathrm{res}}(\hat{\mathbf{r}}, \hat{\mathbf{s}}; \bm{\theta})$. 

The chosen architecture of the neural networks is composed of fully connected layers with $\tanh$ activation functions, except for the last layer where the absolute value function was used to make the output non-negative. The reciprocity was learned implicitly by adding duplicates of the measurements to the data set, while switching the positions referring to source and receiver.

The kernel function with this weighting function represented by a neural network cannot be analytically obtained. We approximate the plane wave expansion with numerical integration by discretizing $\mathbb{S}^2$ as
\begin{align}
&\kappa_{\mathrm{res}}(\mathbf{r}|\mathbf{s}, \mathbf{r}^\prime|\mathbf{s}^\prime; \bm{\theta})\notag \\
& = \overline{\mathcal{T}}\left (w_{\mathrm{res}}(\hat{\mathbf{r}}, \hat{\mathbf{s}};\theta)\frac{\mathrm{e}^{-\mathrm{i}k(\hat{\mathbf{r}}\cdot \mathbf{r}^\prime + \hat{\mathbf{s}}\cdot \mathbf{s}^\prime)} + \mathrm{e}^{-\mathrm{i}k(\hat{\mathbf{r}}\cdot \mathbf{s}^\prime + \hat{\mathbf{s}}\cdot \mathbf{r}^\prime)} }{2}; \mathbf{r}|\mathbf{s} \right ),
\end{align}
where $\overline{\mathcal{T}}$ represents an approximation of $\mathcal{T}$ by numerical integration. Note that the fundamental acoustic properties can still be guaranteed by $\kappa_{\mathrm{res}}$.



\subsection{Optimization of hyperparameters for kernel functions}

Finally, the kernel function is represented as
\begin{align}
    \kappa(\mathbf{r}|\mathbf{s}, \mathbf{r}^\prime|\mathbf{s}^\prime; \bm{\alpha}, \bm{\beta}, \bm{\theta}) = \kappa_{\mathrm{dir}}(\mathbf{r}|\mathbf{s}, \mathbf{r}^\prime|\mathbf{s}^\prime; \bm{\alpha}, \bm{\beta}) + \kappa_{\mathrm{res}}(\mathbf{r}|\mathbf{s}, \mathbf{r}^\prime|\mathbf{s}^\prime; \bm{\theta}),
\end{align}
and the estimate $\hat{h}_{\mathrm{R}}$ can be obtained by kernel ridge regression~\eqref{eq:krr}. The hyperparameters of this kernel function, i.e., $\bm{\alpha}$, $\bm{\beta}$, and $\bm{\theta}$, are obtained by minimizing the leave-one-out cross-validation ($\mathrm{LOO}$) error $E_\mathrm{LOO}$ on the data set $\{(q_n, y_n)\}_{n=1}^N$ defined as
\begin{equation}\label{eq:trainloss}
E_\mathrm{LOO}(\bm{\alpha},\bm{\beta},\bm{\theta}) = \frac{1}{N}\sum_{n=1}^N |\hat{h}_\mathrm{R}(\mathbf{q}_n; \breve{Q}_n, \breve{\mathbf{y}}_n, w) - y_n|^2,
\end{equation}
where $\breve{Q}_n$ represents the set of all source/receiver pairs except $\mathbf{q}_n$, and $\breve{\mathbf{y}}_n$ is the vector of every measurement except $y_n$. This error function $E_{\mathrm{LOO}}$ can be computed in a closed form~\cite{Sellamanickan:NC2001}. The hyperparameters associated with the kernel that minimize $E_{\mathrm{LOO}}$ are obtained by using the gradient descent method for $\bm{\beta}$ and $\bm{\theta}$ and the reduced gradient descent method~\cite{Horiuchi:WASPAA2021, Luenberger:Linear_and_Nonlinear_programming} for $\bm{\alpha}$. The gradients of $E_{\mathrm{LOO}}$ in regards to $\bm{\alpha}$, $\bm{\beta}$, and $\bm{\theta}$ are obtained with automatic differentiation~\cite{Flux.jl-2018}. Knowing $\overline{\mathcal{T}}$ is a bounded linear operator\cite{Rudin:FunctionalAnalysis}, we can efficiently compute the gradients without the need to recalculate $E_\mathrm{LOO}$ for each iteration, and the derivation can be further accelerated by using a GPU.

\section{Numerical experiments}

\begin{figure}[!t]
\centering
\centerline{\includegraphics[width=1.0\columnwidth]{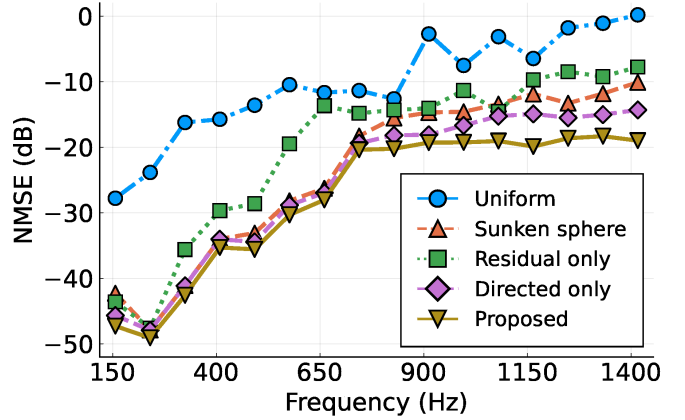}}
  \caption{$\mathrm{NMSE}$ in $\mathrm{dB}$ as a function of frequency.}
  \label{fig:nmse}
\end{figure}

\begin{figure}[!t]
\centering
 \subfloat[\textbf{True}]{\includegraphics[width=0.44\columnwidth]{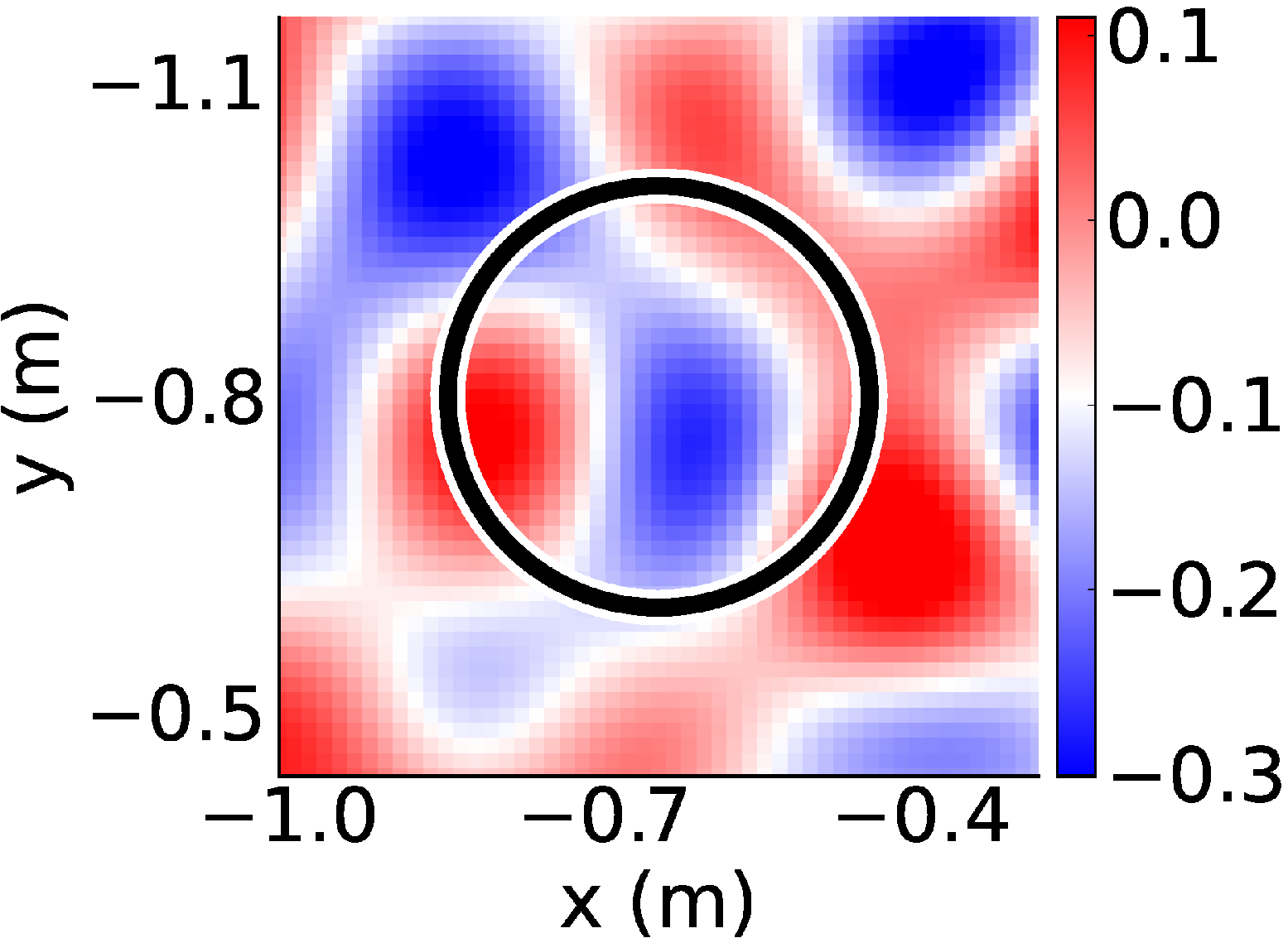}} \ \ \ \
  \subfloat[\textbf{Uniform}]{\includegraphics[width=0.44\columnwidth]{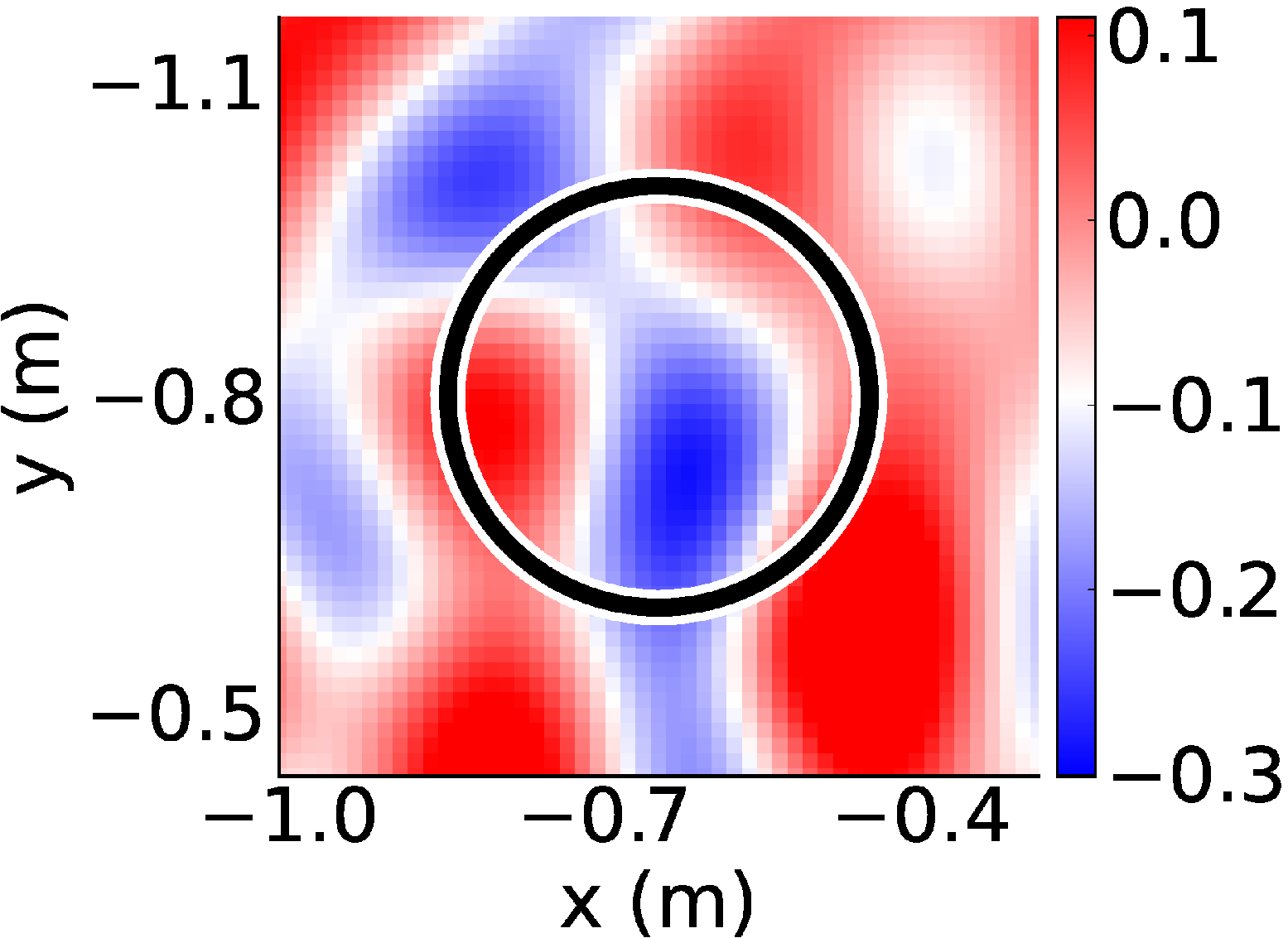}}
  \\
  \centering
   \subfloat[\textbf{Sunken sphere}]{\includegraphics[width=0.44\columnwidth]{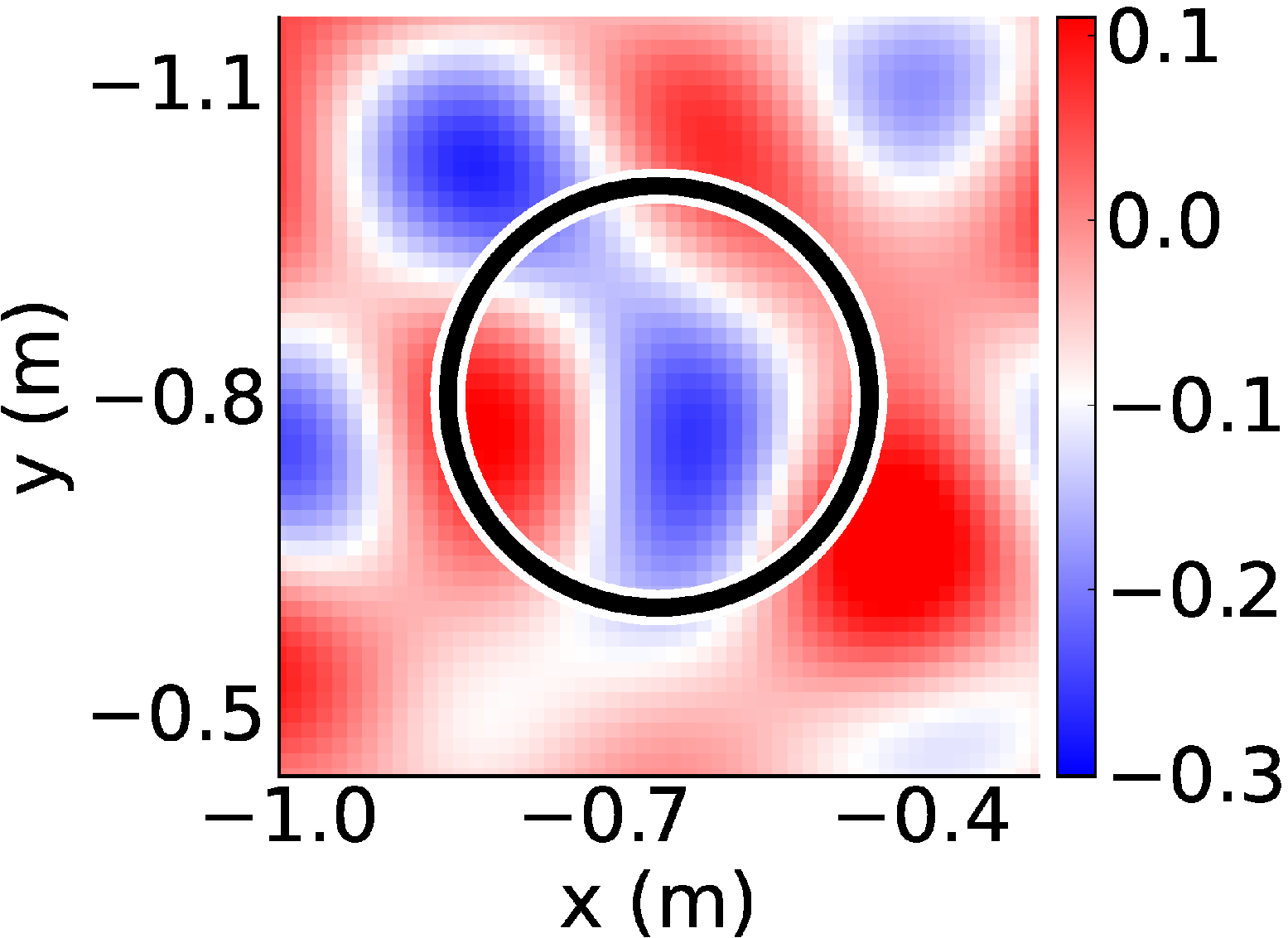}} \ \ \ \ 
     \subfloat[\textbf{Proposed}]{\includegraphics[width=0.44\columnwidth]{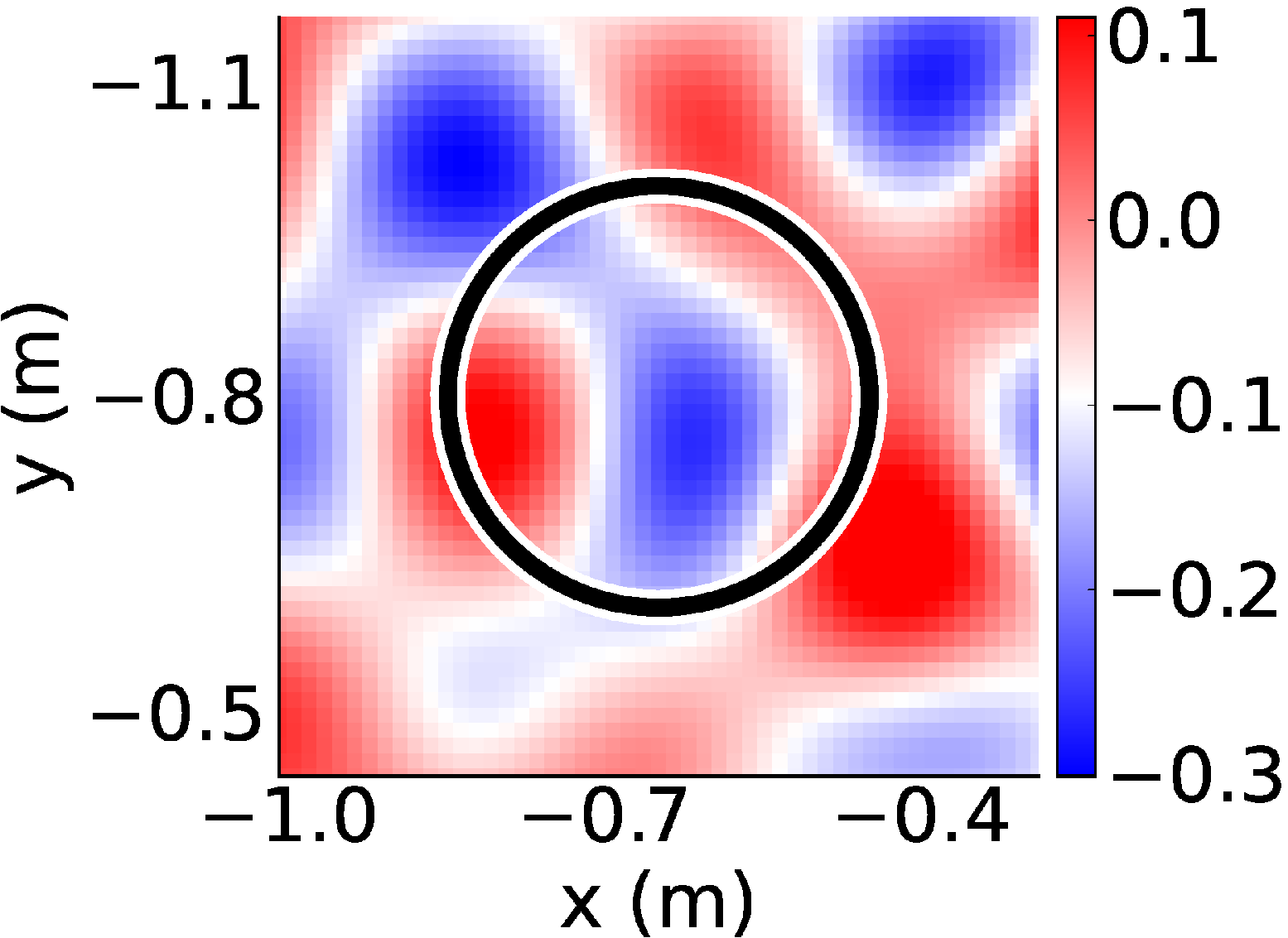}}
  \caption{Distributions of true and estimated ATFs in $\Omega_{\mathrm{R}}$ for the fixed source position $[0.65,\ 0.80,\ 0.48]^\mathsf{T}~\mathrm{m}$ at $1150~\mathrm{Hz}$. Black circles indicate the boundary of $\Omega_\mathrm{R}$.}
  \label{fig:reconst}
  \vspace{10pt}
\centering
  \subfloat[\textbf{Uniform}]{\includegraphics[width=0.44\columnwidth]{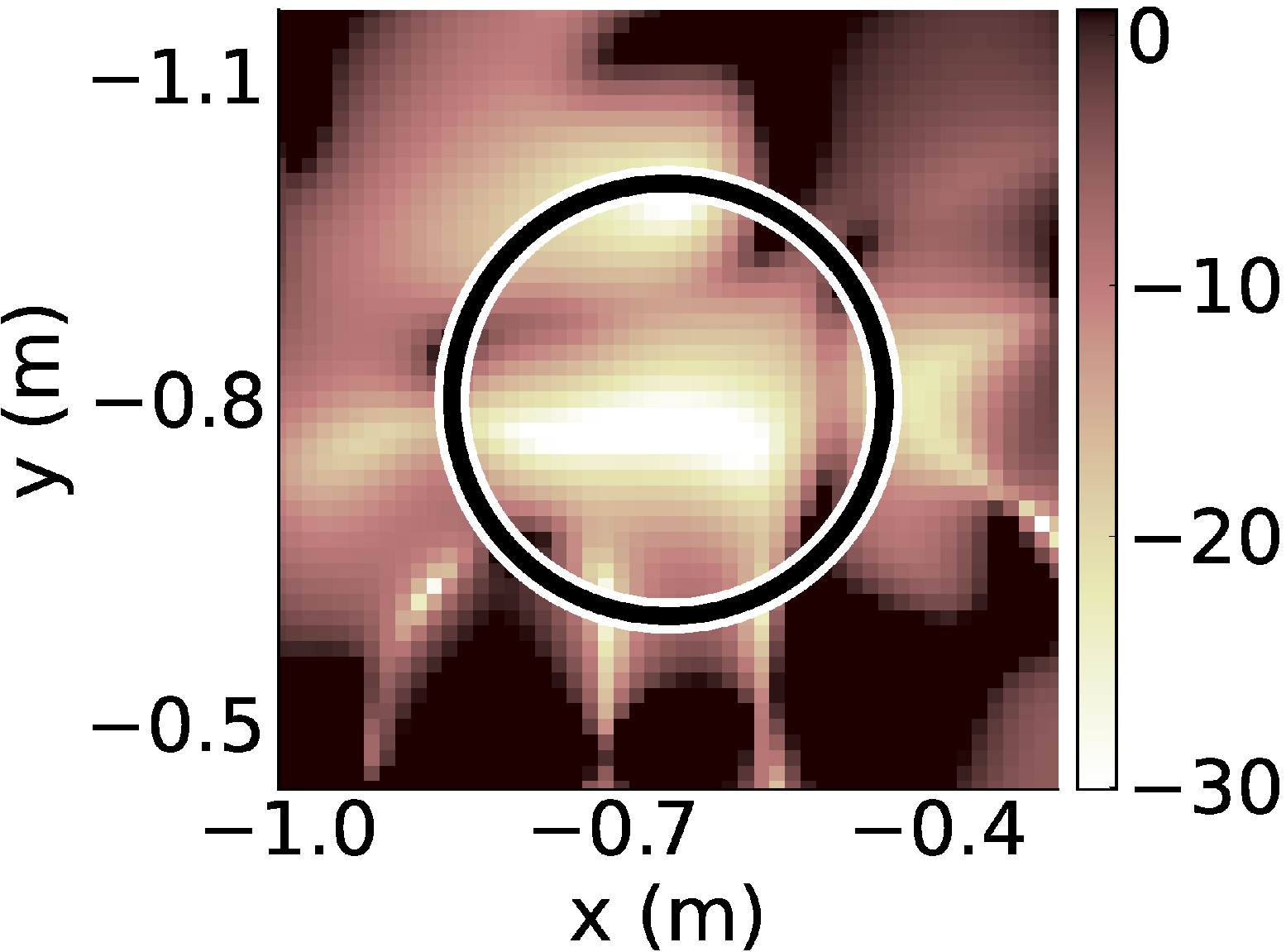}}
  \\
  \centering
  \subfloat[\textbf{Sunken sphere}]{\includegraphics[width=0.44\columnwidth]{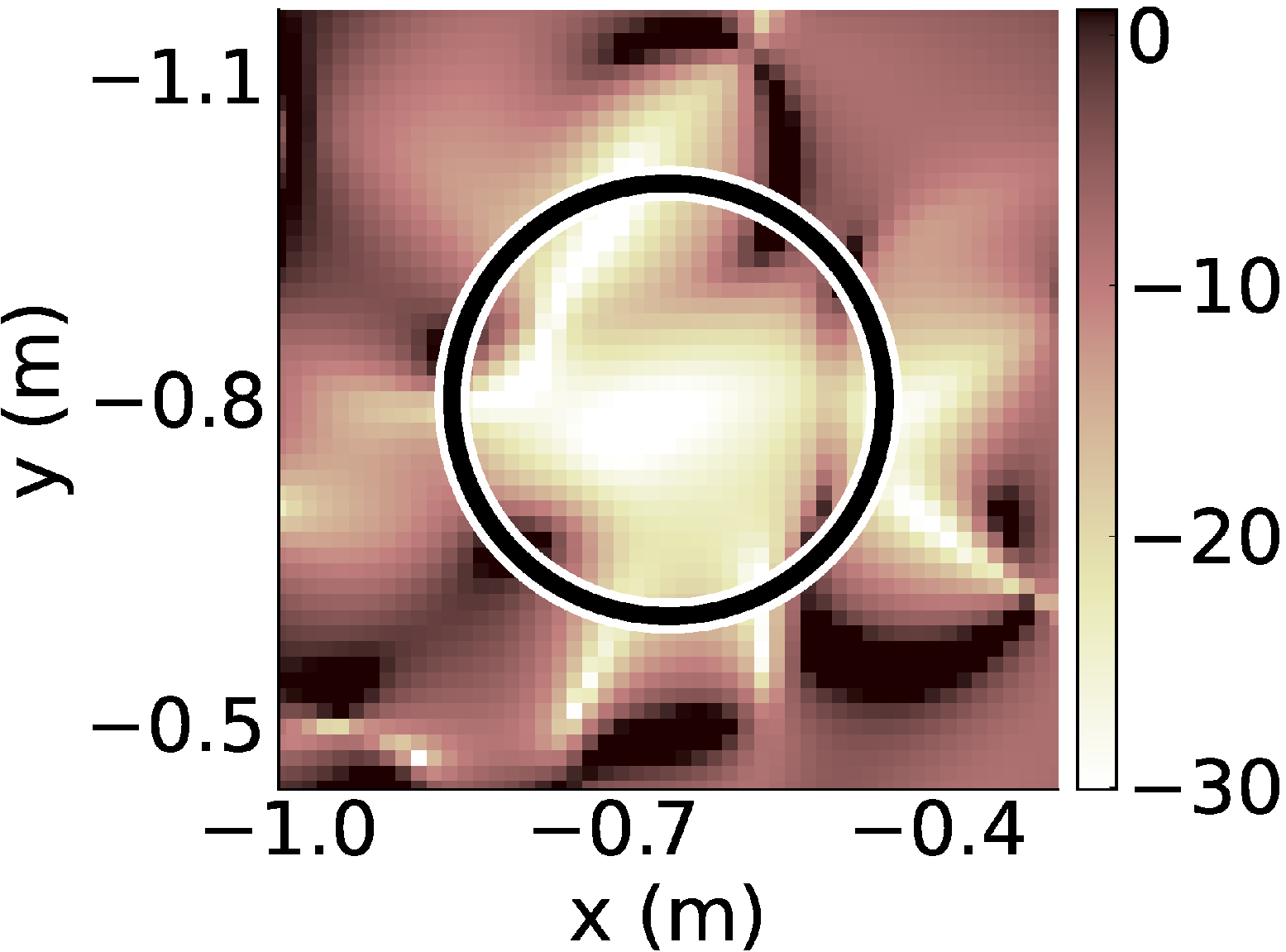}} \ \ \ \ 
  \subfloat[\textbf{Proposed}]{\includegraphics[width=0.44\columnwidth]{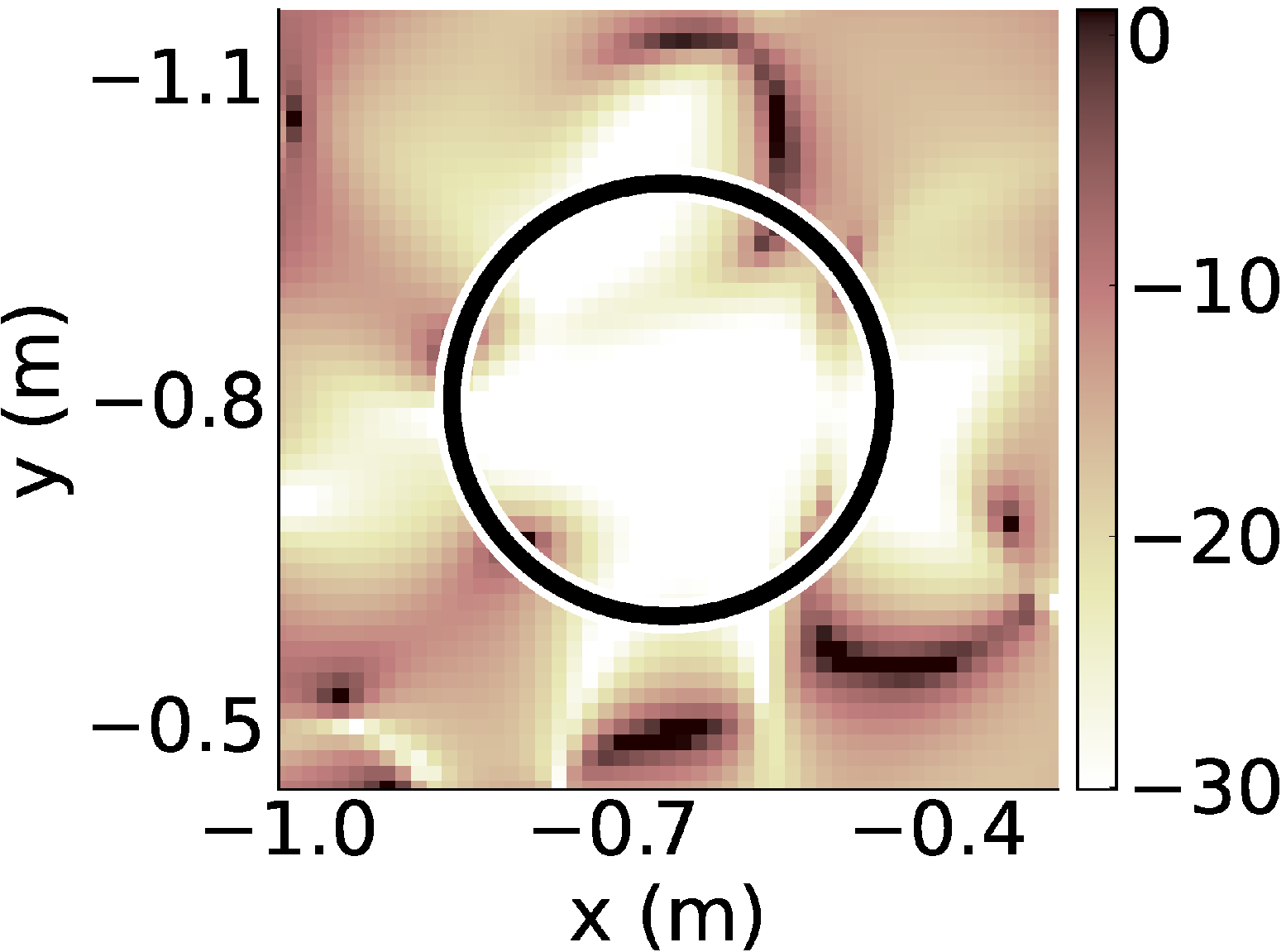}}
  \caption{Normalized square error distribution at $1150~\mathrm{Hz}$ inside $\Omega_{\mathrm{R}}$ in $\mathrm{dB}$.}
  \label{fig:NSE}
\end{figure}

\begin{figure}[!t]
\centering
  \subfloat[Directed reverberation $\varphi_{\mathrm{dir}}$]{\includegraphics[width=0.79\columnwidth]{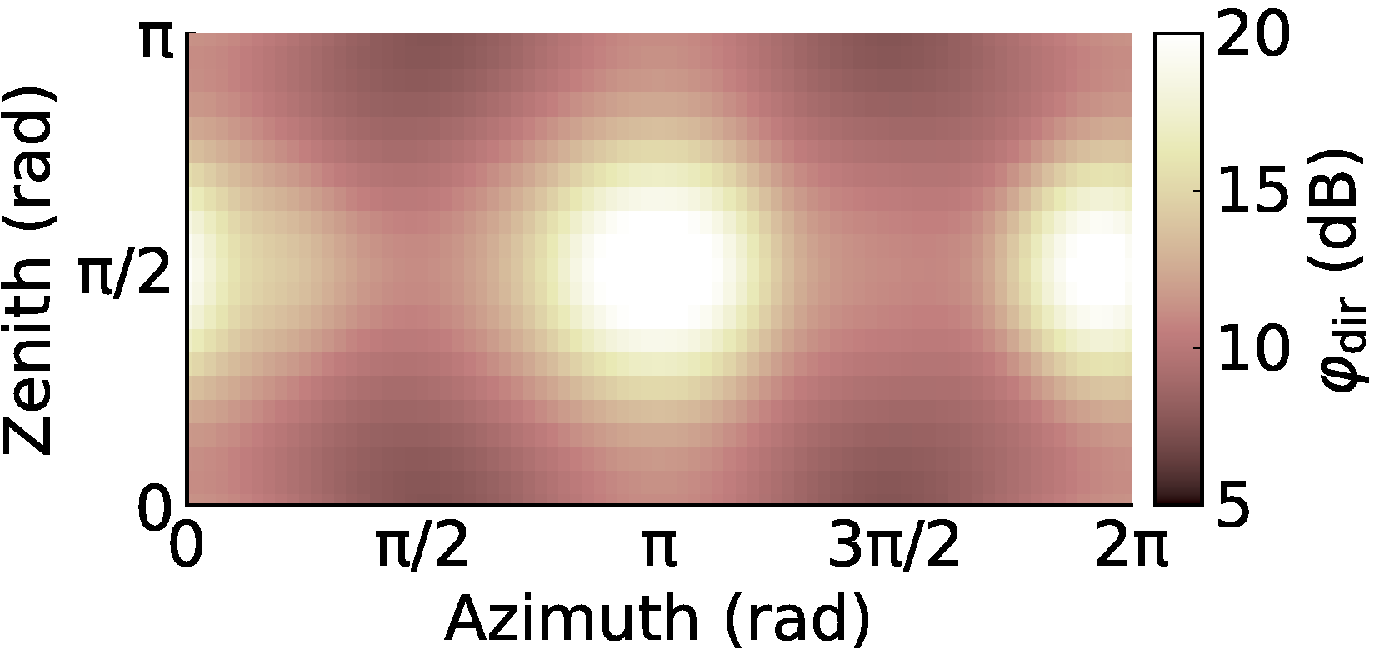}}
  \\
  \centering
  \subfloat[Residual reverberation $w_{\mathrm{res}}$]{\includegraphics[width=0.79\columnwidth]{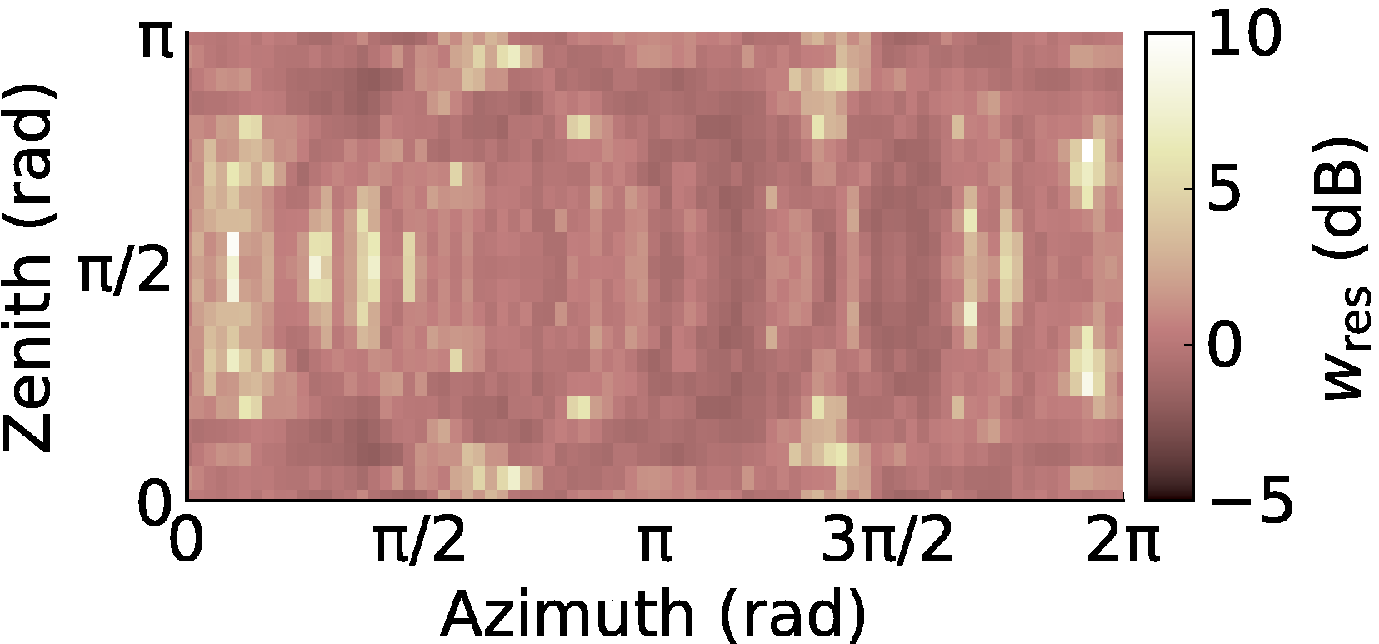}}
  \caption{Directional weighting function in $\mathrm{dB}$ obtained by \textbf{Proposed}. The direction $\hat{\mathbf{s}}$ for $w_{\mathrm{res}}$ was fixed to $[1/\sqrt{3}\ 1/\sqrt{3}\ 1/\sqrt{3}]^\mathsf{T}$.}
  \label{fig:weights}
\end{figure}

We evaluated our proposed method in numerical experiments. The image source method~\cite{Allen:JASA1979} was used to simulate a 3D acoustic environment. We compared several weighting functions for the kernel interpolation of ATFs: uniform weight~\cite{Ribeiro:IEEE_ACM_J_ASLP2022} (\textbf{Uniform}), sunken sphere function~\eqref{eq:weight_sunkensphere} (\textbf{Sunken sphere})~\cite{Ribeiro:ICASSP2022}, directed weighting only (\textbf{Directed only}), residual weighting only (\textbf{Residual only}), and the proposed weighting (\textbf{Proposed}).


We assumed $\Omega$ to be a shoebox-shaped room with dimensions $3.2~\mathrm{m}\times 4.0~\mathrm{m}\times 2.7~\mathrm{m}$. The reflection coefficients of the walls were set so that the reverberation time $T_{60}$ corresponds to $0.45~\mathrm{s}$. The center of the coordinate system was at the center of the room. The source and receiver regions were represented by two spheres of $0.2~\mathrm{m}$ radii centered in $\mathbf{s}_0 = [0.65,\ 0.80,\ 0.48]^{\mathsf{T}}~\mathrm{m}$ and $\mathbf{r}_0 = -\mathbf{s}_0$, respectively. The direction connecting the centers was $\hat{\mathbf{v}}_0 = [-0.57,\  -0.70, \ -0.42]^\mathsf{T}$. 


For the derivation of the interpolation functions, the loudspeakers and microphones were arranged on the two spherical layers of $0.2~\mathrm{m}$ and $0.19~\mathrm{m}$ radii on $\Omega_{\mathrm{S}}$ and $\Omega_{\mathrm{R}}$ where $L=M=41$. The positions of the loudspeakers and microphones were set using the spherical $t$-design~\cite{Chen:SIAM2006} for $t=4$ in the outer layer and $t=3$ in the inner layer. We added Gaussian noise to the ATF measurements so that the signal-to-noise ratio ($\mathrm{SNR}$) would become $20~\mathrm{dB}$. 

For \textbf{Sunken sphere}, $\gamma$ and $\beta$ in \eqref{eq:weight_sunkensphere} were also obtained by minimizing $E_\mathrm{LOO}$. For the directed field model in \textbf{Directed only} and \textbf{Proposed}, the components $\{ \hat{\mathbf{v}}_d \}_{d=1}^D$ in \eqref{eq:weight_early_phi_w} were determined using the spherical $t$-design with $t=4$; therefore, the total number of plane waves was $D=25$. For \textbf{Residual only} and $\textbf{Proposed}$, the numerical integration was performed using Lebedev quadrature~\cite{Lebedev:Doklady1999} with 110 points. The architecture of the neural networks was composed of two fully connected hidden layers with 20 neurons each. The regularization parameter $\lambda$ was set to $10^{-3}$ for all the methods.


For the evaluation measure, we define the normalized mean square error (NMSE) as
\begin{equation}
\mathrm{NMSE}(\hat{h}, h) = 10\log_{10} \left ( \frac{\sum_{n=1}^{N^\prime} \left | \hat{h}(\mathbf{q}^\prime_n) - h(\mathbf{q}^\prime_n) \right |^2}{\sum_{n=1}^{N^\prime} \left | h(\mathbf{q}^\prime_n) \right |^2}\right ),
\end{equation}
where $\{\mathbf{q}^\prime_{n}\}_{n=1}^{N^\prime}$ are the evaluation points, sampled randomly within the regions. The total number of evaluation points was $N^\prime=M^\prime L^\prime=9025$ with $L^\prime=M^\prime=95$. 


Fig.~\ref{fig:nmse} shows the NMSE. \textbf{Directed only} and \textbf{Proposed} performed best for most frequencies, showing that a higher degree of adaptability improves estimations. The inclusion of the residual field model also improved the frequency performance of \textbf{Proposed} compared to \textbf{Directed only}. As an example, the ATF distribution and normalized square error distributions in $\Omega_{\mathrm{R}}$ for the fixed source position $\mathbf{s}_0=[0.65,\ 0.80,\ 0.48]^{\mathsf{T}}~\mathrm{m}$ (the center of $\Omega_\mathrm{S}$) at $1150~\mathrm{Hz}$ are shown for \textbf{Uniform}, \textbf{Sunken sphere}, and \textbf{Proposed} in Fig.~\ref{fig:NSE}. $\textbf{Proposed}$ could reconstruct the sound field more accurately and in a larger effective area than the established methods, demonstrating smaller error for most points within $\Omega_\mathrm{R}$. The directional weighting function obtained by the proposed method for $\hat{\mathbf{s}}=[1/\sqrt{3}, 1/\sqrt{3}, 1/\sqrt{3}]^{\mathsf{T}}$ is shown in Fig.~\ref{fig:weights}. The plot of $\varphi_\mathrm{dir}$ displays greater amplitudes than $w_\mathrm{res}$, concentrated in a few directions, whereas the structure of $w_{\mathrm{res}}$ is more erratic, with smaller amplitudes densely assigned, as was expected. The performance of \textbf{Proposed} shows that modeling the directed and residual fields separately provides a more complete model for the ATF that can deliver reliable approximations in frequency-wise and position-wise bases.



\section{Conclusion}
We proposed a region-to-region ATF interpolation method based on kernel interpolation with an adaptive kernel for directed and residual reverberations. Whereas the current region-to-region ATF interpolation methods do not consider acoustic properties for which measurements are performed, our proposed method adapts to the acoustic environment by optimizing the directional weighting function related to the kernel function. The directed field is approximated by superposition of unimodal functions, and the residual field is modeled using neural networks. Their hyperparameters are optimized by minimizing the cross validation error. The effectiveness of the proposed method compared with the current methods was validated, in particular at high frequencies, by numerical experiments. 

\section{Acknowledgements}
This work was supported by JSPS KAKENHI
under Grant number 19H01116, 22H03608, and JST FOREST Program under Grant number JPMJFR216M.

\bibliographystyle{IEEEbib_mod}
\bibliography{str_def_abrv, koyama_en, refs}

\begin{thebibliography}{10}

\bibitem{IIR_ATF}
Y.~Haneda, S.~Makino, Y.~Kaneda, and N.~Koizumi,
\newblock ``{ARMA} modeling of a room transfer function at low frequencies,''
\newblock {\em J. Acoust. Soc. Japan (E)}, vol. 15, pp. 353--355, 1994.

\bibitem{Haneda:IEEE_J_SAP1999}
Y.~Haneda, Y.~Kaneda, and N.~Kitawaki,
\newblock ``Common-acoustical-pole and residue model and its application to
  spatial interpolation and extrapolation of a room transfer function,''
\newblock {\em {IEEE} Trans. Speech Audio Process.}, vol. 7, no. 6, pp.
  709--717, 1999.

\bibitem{Mignot:IEEE_ACM_J_ASLP2014}
R.~Mignot, G.~Chardon, and L.~Daudet,
\newblock ``Low frequency interpolation of room impulse responses using
  compressed sensing,''
\newblock {\em {IEEE/ACM} Trans. Audio, Speech, Lang. Process.}, vol. 22, no.
  1, pp. 205--216, 2014.

\bibitem{Antonello:TASLP2017}
N.~Antonello, E.~De~Sena, M.~Moonen, P.~A. Naylor, and T.~van Waterschoot,
\newblock ``Room impulse response interpolation using a sparse spatio-temporal
  representation of the sound field,''
\newblock {\em {IEEE/ACM} Trans. Audio, Speech, Lang. Process.}, vol. 25, no.
  10, pp. 1929--1941, 2017.

\bibitem{Das:ICASSP2021}
O.~Das, P.~Calamia, and S.~V.~A. Gari,
\newblock ``Room impulse response interpolation from a sparse set of
  measurements using a modal architecture,''
\newblock in {\em Proc. {IEEE} Int. Conf. Acoust., Speech, Signal Process.
  ({ICASSP})}, 2021, pp. 960--964.

\bibitem{Pezzoli:Sensors2022}
M.~Pezzoli, D.~Perini, A.~Bernardini, F.~Borra, F.~Antonacci, and A.~Sarti,
\newblock ``Deep prior approach for room impulse response reconstruction,''
\newblock {\em Sensors}, vol. 22, no. 7, 2710, 2022.

\bibitem{Samarasinghe}
P.~N. {Samarasinghe}, T.~D. {Abhayapala}, M.~A. {Poletti}, and T.~{Betlehem},
\newblock ``An efficient parameterization of the room transfer function,''
\newblock {\em {IEEE/ACM} Trans. Audio, Speech, Lang. Process.}, vol. 23, no.
  12, pp. 2217--2227, 2015.

\bibitem{Ribeiro:IEEE_SAM2020}
J.~G.~C. Ribeiro, N.~Ueno, S.~Koyama, and H.~Saruwatari,
\newblock ``Kernel interpolation of acoustic transfer function between regions
  considering reciprocity,''
\newblock in {\em Proc. {IEEE} Sensor Array Multichannel Signal Process.
  Workshop ({SAM})}, 2020.

\bibitem{Ribeiro:IEEE_ACM_J_ASLP2022}
J.~G.~C. Ribeiro, N.~Ueno, S.~Koyama, and H.~Saruwatari,
\newblock ``Region-to-region kernel interpolation of acoustic transfer
  functions constrained by physical properties,''
\newblock {\em {IEEE/ACM} Trans. Audio, Speech, Lang. Process.}, vol. 30, pp.
  2944--2954, 2022.

\bibitem{Ribeiro:ICASSP2022}
J.~G.~C. Ribeiro, S.~Koyama, and H.~Saruwatari,
\newblock ``Region-to-region kernel interpolation of acoustic transfer function
  with directional weighting,''
\newblock in {\em Proc. {IEEE} Int. Conf. Acoust., Speech, Signal Process.
  ({ICASSP})}, Singapore, 2022, pp. 576--580.

\bibitem{Horiuchi:WASPAA2021}
R.~Horiuchi, S.~Koyama, J.~G.~C. Ribeiro, N.~Ueno, and H.~Saruwatari,
\newblock ``Kernel learning for sound field estimation with {L}1 and {L}2
  regularizations,''
\newblock in {\em Proc. {IEEE} Int. Workshop Appl. Signal Process. Audio
  Acoust. ({WASPAA})}, 2021, pp. 261--265.

\bibitem{Shigemi:IWAENC2022}
K.~Shigemi, S.~Koyama, T.~Nakamura, and H.~Saruwatari,
\newblock ``Physics-informed convolutional neural network with bicubic spline
  interpolation for sound field estimation,''
\newblock in {\em Proc. Int. Workshop Acoust. Signal Enhancement ({IWAENC})},
  2022.

\bibitem{Colton2003}
D.~Colton and P.~Monk,
\newblock ``{Herglotz Wave Functions in Inverse Electromagnetic Scattering
  Theory},''
\newblock in {\em Topics in Computational Wave Propagation: Direct and Inverse
  Problems}, M.~Ainsworth, P.~Davies, D.~Duncan, B.~Rynne, and P.~Martin, Eds.,
  pp. 367--394. Springer, Berlin, 2003.

\bibitem{Ikehata:HMJ2005}
M.~Ikehata,
\newblock ``{The Herglotz wave function, the Vekua transform and the enclosure
  method},''
\newblock {\em Hiroshima Math. J.}, vol. 35, 2005.

\bibitem{Murphy:ProbML}
K.~P. Murphy,
\newblock {\em Probabilistic Machine Learning},
\newblock MIT Press, Massachusetts, 2022.

\bibitem{Mardia:Directional_Statistics}
K.~V. Mardia and P.~E. Jupp,
\newblock {\em Directional Statistics}, vol. 494,
\newblock John Wiley \& Sons, Hoboken, 2009.

\bibitem{Sellamanickan:NC2001}
S.~Sundararajan and S.~Keerthi,
\newblock ``Predictive approaches for choosing hyperparameters in {G}aussian
  processes,''
\newblock {\em Neural Comput.}, vol. 13, pp. 1103--18, 2001.

\bibitem{Luenberger:Linear_and_Nonlinear_programming}
D.~G. Luenberger and Y.~Ye,
\newblock {\em Linear and Nonlinear Programming}, vol. 228,
\newblock Springer Cham, Gewerbestrasse, 2016.

\bibitem{Flux.jl-2018}
M.~Innes, E.~Saba, K.~Fischer, D.~Gandhi, M.~C. Rudilosso, N.~M. Joy,
  T.~Karmali, A.~Pal, and V.~Shah,
\newblock ``Fashionable modelling with flux,''
\newblock {\em Comput. Res. Repo. {(CoRR)}}, 2018.

\bibitem{Rudin:FunctionalAnalysis}
W.~Rudin,
\newblock {\em Functional Analysis},
\newblock McGraw-Hill, New York City, 1991.

\bibitem{Allen:JASA1979}
J.~B. Allen and D.~A. Berkley,
\newblock ``Image method for efficiently simulating small-room acoustics,''
\newblock {\em J. Acoust. Soc. Amer.}, vol. 65, no. 4, pp. 943--950, 1979.

\bibitem{Chen:SIAM2006}
X.~Chen and R.~S. {Womersley},
\newblock ``Existence of solutions to systems of underdetermined equations and
  spherical designs,''
\newblock {\em SIAM J. Numer. Anal.}, vol. 44, no. 6, pp. 2326--2341, 2006.

\bibitem{Lebedev:Doklady1999}
V.~I. Lebedev and D.~N. Laikov,
\newblock ``A quadrature formula for the sphere of the 131st algebraic order of
  accuracy,''
\newblock {\em Doklady Math.}, vol. 59, pp. 477--481, 1999.

\end{thebibliography}

\end{document}